\documentclass[runningheads]{cl2emult}

\usepackage{makeidx}  
\usepackage{graphicx} 
\usepackage{subeqnar} 
\usepackage{multicol} 
\usepackage{cropmark} 
\usepackage{lnp}      
\makeindex            

\newcommand{\euler}[1]{{\usefont{U}{eur}{m}{n}#1}}

\newcommand{\umu}{\mbox{\euler{\char22}}}

\begin{document}
\title*{Infrared Properties of High Redshift and X-ray Selected AGN Samples}
\toctitle{Infrared Properties of High Redshift and X-ray Selected AGN Samples}
%
%
\titlerunning{IR Properties of High-$z$ and X-ray Selected AGN}
%
\author{Belinda J. Wilkes\inst{1}
\and Eric J. Hooper\inst{1}
\and Kim K. McLeod\inst{2}
\and Martin S. Elvis\inst{1}
\and David H. Hughes\inst{3}
\and Chris D. Impey\inst{4}
\and Joanna K. Kuraszkiewicz\inst{1}
\and Carol S. Lonsdale\inst{5}
\and Matt A. Malkan\inst{6}
\and Jonathan C. McDowell\inst{1}
}

\authorrunning{Belinda J. Wilkes et al.}
%
%
\institute{
Harvard-Smithsonian CfA, Cambridge, MA 02138 USA
\and Wellesley College, Wellesley, MA 02481 USA
\and INAOE, Puebla, Pue. Mexico
\and Steward Observatory, Tucson, AZ 85721 USA
\and IPAC, Caltech, Pasadena, CA 91125 USA
\and UCLA, Los Angeles, CA 90095-1562 USA 
}

\maketitle              

\begin{abstract}
The NASA/ISO Key Project on active galactic nuclei (AGN) seeks to better
understand the broad-band spectral energy distributions (SEDs) of these
sources from radio to X-rays, with particular emphasis on infrared
properties.  The ISO sample includes a wide variety of AGN types and
spans a large redshift range.  Two subsamples are considered herein:
8 high-redshift ($1 < z < 4.7$) quasars; and 22 hard X-ray selected
sources.  

The X-ray selected AGN show a wide range of IR continuum shapes, extending
to cooler colors than the optical/radio sample of~\cite{Elvis94}.
Where a far-IR turnover is clearly observed, the slopes are $< 2.5$ in
all but one case so that non-thermal emission remains a possibility.
The highest redshift quasars show extremely strong, hot IR continua
requiring $\sim 100 M_{\odot}$ of 500--1000$\ts K$ dust with $\sim 100$
times weaker optical emission. Possible explanations for these unusual
properties include: reflection of the optical light from material
above/below a torus; strong obscuration of the optical continuum; or 
an intrinsic deficit of optical emission.  A cosmology of ($H_0$,
${\Omega}_m$, ${\Omega}_k$, ${\Omega}_{\Lambda}$) = (50 km s$^{-1}$
Mpc$^{-1}$, 1, 0, 0) is assumed.

\end{abstract}

\section{Introduction}
%

Active galactic nuclei\index{active galactic nuclei} are among
the broadest emission sources in nature, producing significant flux
over a span 
$> 9$ decades in frequency, from radio to X-rays and
beyond~\cite{Elvis94}.  The various emission mechanisms involved are
presumably ultimately powered by a central supermassive black
hole~\cite{Rees84}.

A substantial fraction of the bolometric luminosity of many AGN
emerges in the infrared, from synchrotron radiation\index{synchrotron
radiation} and dust\index{dust emission}.  Which of these is the
principal emission mechanism is related to quasar type and is an open
question in many cases~\cite{Wilkes99a}.  Non-thermal emission is
paramount in core dominated radio-loud quasars and
blazars~\cite{Impey88},
although hot dust contributes in some cases~\cite{Courvoisier98}.  The
non-thermal component is likely related to radio and higher
frequency synchrotron radiation,
providing information about the relativistic plasma and magnetic
fields associated with quasars.  Other AGN classes show evidence for a
predominant dust contribution~\cite{Edelson87}, particularly
infrared-luminous radio-quiet quasars~\cite{Hughes93}, or a mix of
emission components~\cite{Haas98}.
Much of the dust emission is due to heating by higher energy photons
from the active nucleus,
and is therefore important for understanding the overall energy
balance.
The AGN thermal component may be an orientation-independent parameter,
useful for examining unification hypotheses.
%

The nature of the foremost infrared emission source is ambiguous in
many AGN with sparsely sampled spectral energy
distributions\index{spectral energy distribution} (SEDs).
Dust with smooth spatial and temperature distributions can mimic a
power law spectrum~\cite{Rees69,Bollea76}, particularly in the absence
of detailed measurements to reveal bumps from temperature and density
inhomogeneities.  Grain emissivity is characterized by a Planck
function multiplied by a power law factor $\propto
\nu^{1 - 2}$~\cite{Hildebrand83}, so spectral slopes in the
Rayleigh-Jeans region of the coldest potential thermal component lie
between $\alpha = 2$--4, depending on the grain properties and
optical depth.  Known synchrotron emitters have relatively flat sub-mm
power-law spectra ($\alpha \leq 1.1; f_{\nu} \propto
\nu^{\alpha}$)~\cite{Gear94}, and radio sources generally have spectra
flatter than the canonical $\alpha = 2.5$ for a self-absorbed
homogeneous synchrotron source~\cite{ODea98}.  Therefore, $\alpha = 2.5$
is a convenient partition to distinguish thermal emission from
standard non-thermal models, and simple two-point spectral slopes and
even lower limits to spectral indices may reveal the dominant
mechanism.  However, synchrotron models with a concave electron energy
distribution~\cite{deKool89,Schlickeiser91}, free-free absorption, or
plasma suppression~\cite{Schlickeiser89} can produce slopes steeper
than $\alpha = 2.5$.  While $\alpha = 4$ is observed in some
milliarcsecond radio knots~\cite{Matveyenko99}, thermal models offer
the most consistent explanation for steep far-infrared (FIR) to mm
slopes~\cite{Hughes93,Andreani99}.  A thermal origin is considered to
be the most likely explanation for sub-mm/FIR slopes $\alpha > 2.5$ in
the present work.


Two large, complementary ISO AGN observing programs are opening up new
wavelength windows in the FIR as well as improving the spatial resolution and
sampling at shorter wavelengths: the ISO European Central Quasar
Program~\cite{Haas98,Haas99}; and the NASA/ISO Key Project on AGN, discussed
herein.  
The Key Project sample consists of 5--200$\ts \umu \rm{m}$ chopped and
rastered ISOPHOT~\cite{Lemke96} observations of 73 AGN selected to
incorporate a wide range of AGN types and redshifts.  The data are
reduced using a combination of the (ISO-) PHOT Interactive Analysis
(PIA)~\cite{Gabriel98} software plus custom scripts.  Details of the
reduction, difficulties encountered, and early results are discussed
in~\cite{Wilkes97,Hooper99a,Wilkes99b,Hooper99b}.  
These observations directly measure the FIR spectral slopes in low and
moderate redshift AGN and provide better constraints on the emission
mechanisms throughout the infrared region.  Some of the fundamental
questions being addressed with the new data include: the range of SEDs
within each quasar type; differences between one type and another; the
evolution of the SEDs; and correlations with fluxes at other
\nopagebreak
wavebands, host galaxy properties, and orientation indicators.  
\nopagebreak
This paper focuses on two subsamples, hard X-ray selected AGN, and
high-redshift quasars.

\section{Hard X-ray Selected AGN}

Infrared and X-ray data complement each other well and are important
for understanding the overall AGN energy balance.  Non-thermal
infrared emission is possibly connected with the X-rays, either
directly as a portion of a broad synchrotron component, 
or as part of a radio-infrared seed spectrum which Compton scatters to
produce the X-rays.
Infrared data from dust-dominated sources
reveal the level of ultra-violet (UV) and soft X-ray radiation which has
been reprocessed, and dust masses can be estimated assuming optically thin
emission.

\begin{figure}[h]
\centering
\vskip -0.2in
\includegraphics[width=.8\textwidth]{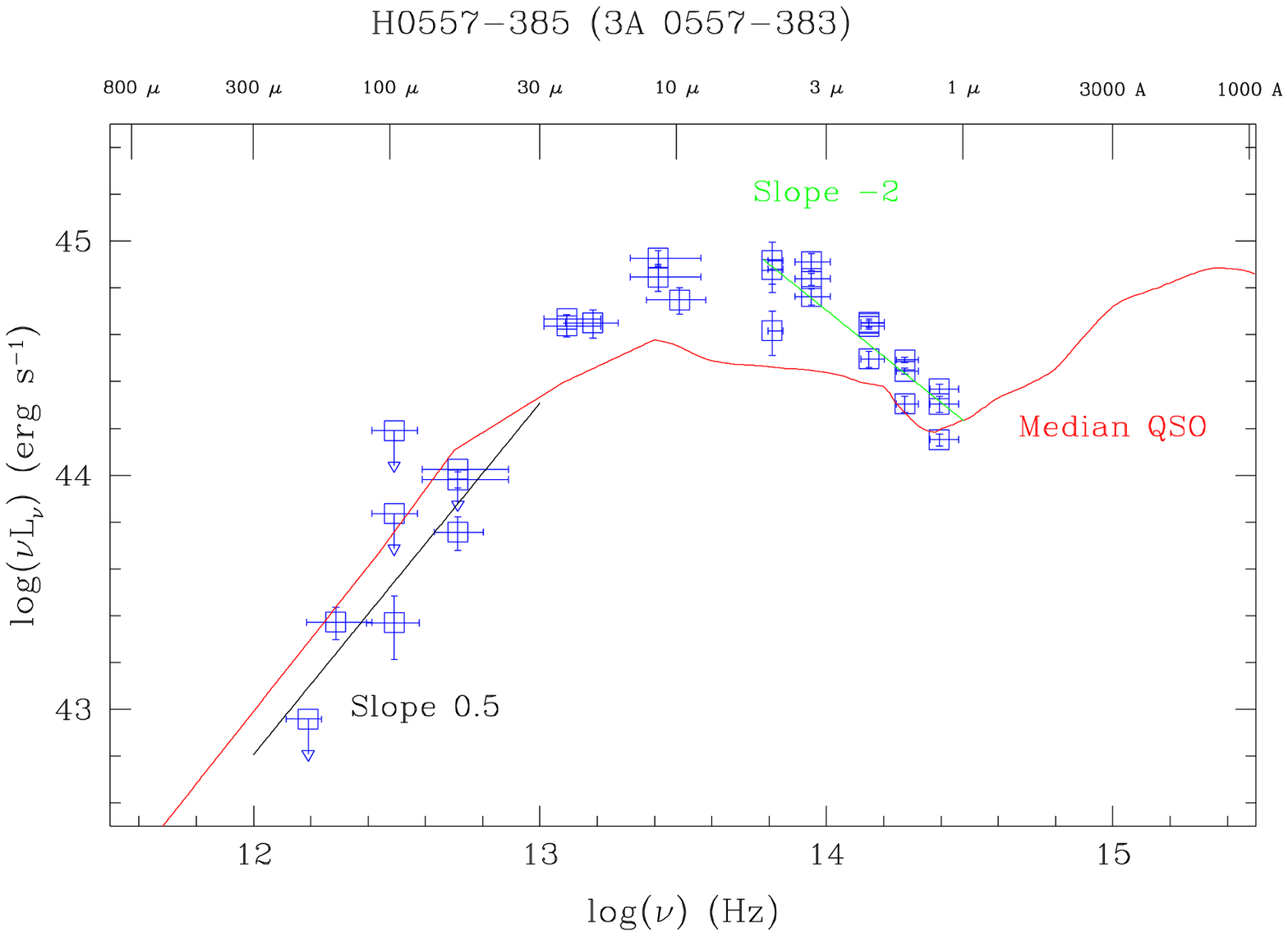}
\vskip -1.3in
\includegraphics[width=.8\textwidth]{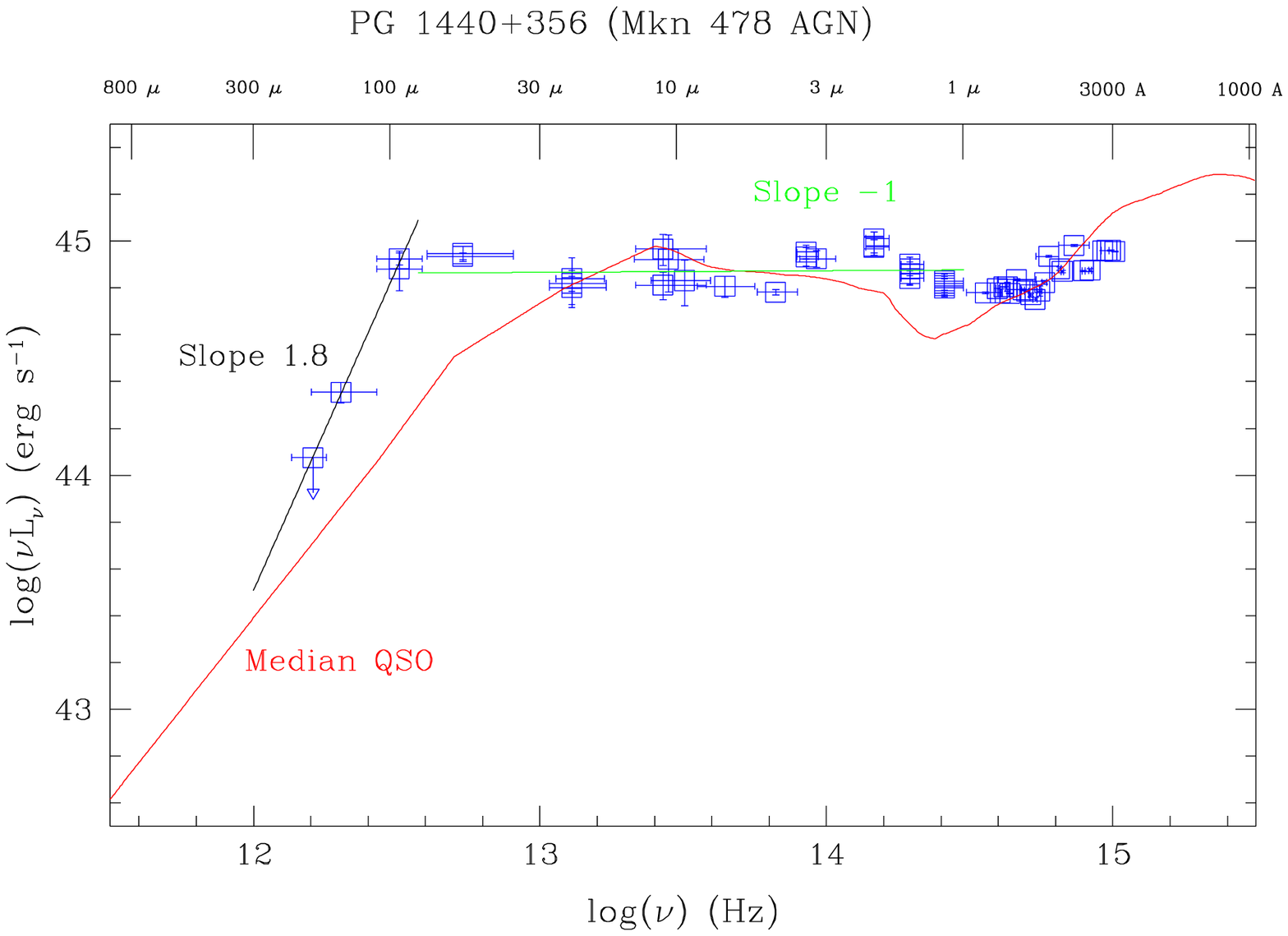}
\vskip -1.2in
\caption{\it Spectral Energy Distributions (SEDs) of H0557-385 (top) and PG1440+356 (bottom) 
compared with the median SED for a low-redshift sample~\cite{Elvis94}
illustrating the two extreme SED types present in this sample.
The near-IR and far-IR cutoff slopes are indicated in both cases.}
\label{fg:xrayseds}
\vspace*{-0.2in}
\end{figure}

\begin{table}[h]
\centering
\caption{X-ray Selected Sample: IR SED Parameters.}
\label{tb:lums}
\begin{tabular}{llllll}
\hline
Name \hspace*{0.6in} & $z$ \hspace*{0.4in} & L$_{10-100\mu}$ \hspace*{0.2in}
& L$_{1-10\mu}$ \hspace*{0.2in} & $\alpha_{IR}$\hspace*{0.4in} & $\alpha_{cut}$ \hspace*{0.1in}\\
& & $10^{44}$ erg s$^{-1}$ & $10^{44}$ erg s$^{-1}$ \hspace*{0.2in} & & \\
\hline
MKN 1152 & 0.052 & 6.3 & 2.4 & $-$0.9 & $-$\\
MKN 590 &0.025 & 3.3 & 4.2 & $-$0.9 & $-$\\
H0235$-$52 & 0.045 & 4.5 & 2.6 &$-$1.6 & 0.2\\
H0557$-$385 & 0.034 & 11.9 & 3.8 & $-$2.1 & 0.7\\
PG0804+761 & 0.100 & 37.1 & 53.0 & $-$1.9 & 0.6\\
H1039$-$074 & 0.674 & 450. & 190. &$-$ & $-$ \\
NGC 3783 & 0.009 & 1.3 & 1.6 & $-$1.4 & 1.4 \\
TON 1542 & 0.064 & 8.3 & 15.9 &$-$1.7 & $-$\\
IR1321+058 & 0.201 & 210.  & 6.4 &$-$2.8 & 1.8 \\
MCG-6-30-15 & 0.008 & 0.6& 0.4 & $-$1.6 & 1.3 \\
IC4329A & 0.014 & 5.4 & 20.3 & $-$  0.8\\
H1419+480 & 0.072 & 15.5  & 7.2 & $-$1.3 & 0.3\\
PG1440+356 & 0.077 & 18.1 & 17.5 & $-1$.1 & 1.8\\
H1537+339 & 0.330 & 72.3 & 42.7 &$-$1.1 & $-$\\
KAZ1803+676 & 0.136 & 20.9 & 31.0 & $-$1.3 & 0.2\\
E1821+643 & 0.297 & 720. & 550. & $-$1.1 & 2.5\\
H1834$-$653 & 0.013 & 1.2  & 0.3 & $-$1.9 & 1.6\\
MKN 509 & 0.035 & 11.4 & 20.3 & $-$1.0 & 2.2\\
NGC 7213 & 0.006 & 1.0 & 0.6 &$-$ & $-$ \\
MR2251$-$178 & 0.068 & 18.4 & 23.6 & $-$0.9 & 3.7\\
MCG-2-58-22 & 0.048 & 13.0 &10.0 & $-$0.9 & 2.5\\
\hline
\end{tabular}
\vskip -0.2in
\end{table}

Most X-ray selected AGN to date have been observed in soft energy
bands $< 3.5$ keV~\cite{Stocke91,Thomas98,Hasinger98}.  These surveys
suffer obscuration biases similar to optical selection, due to the gas
typically associated with dust.  
The absorption cross-section drops steeply with increasing
energy~\cite{Zombeck90}, so hard X-ray selection is much less affected
by intervening material.  Surveys in hard X-rays arguably are the most
efficient way to distinguish between accreting and stellar
sources~\cite{Fiore99} as well as the optimal method for defining a
representative sample of AGN~\cite{Wilkes99a}.  A comparison between
the UV/soft X-ray flux absorbed and the IR emission
provides an estimate of the relative importance of accretion and
stellar power in AGN.  
We randomly selected 23 ISO targets (of which 22 are reported in
Table~\ref{tb:lums}) from a 2--10 keV AGN sample derived from the A2
experiment onboard the HEAO 1 satellite~\cite{Grossan92,Jahoda89}; 12
of these are also in the earlier Piccinotti
sample~\cite{Kotilainen92,Malizia99,Piccinotti82} from the same
experiment.

\begin{figure}[h]
\centering
\vskip -0.5in
\includegraphics[width=.8\textwidth,angle=-90]{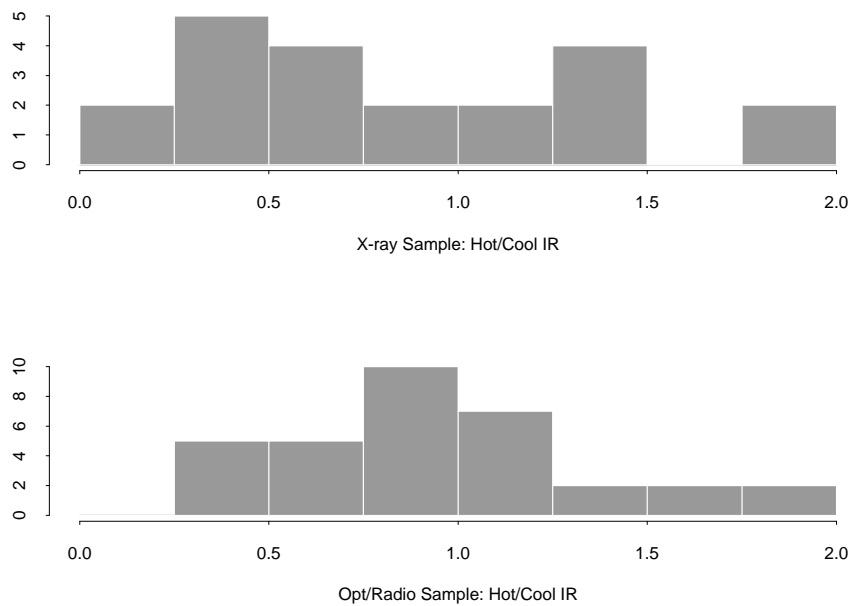}
\vskip -0.1in
\caption{\it A comparison of the distribution of the ratio of hot to cool IR luminosities
(L$(1-10\mu)$/L$(10-100\mu)$)
for the HEAO sample with that of~\cite{Elvis94}. While a suggestive
trend toward stronger cool luminosities is present in the HEAO sample,
a KS test gives only an 18\% probability that this is significant.
}
\label{fg:hotnesshist}
\vspace*{-0.2in}
\end{figure}

Figure~\ref{fg:xrayseds} shows the IR$-$optical SEDs for two HEAO AGN
illustrating the wide range of behavior present.  H0557$-$385 shows a
strong hot component with a steep near-IR slope ($\alpha_{ir} \sim
-2$, F$_{\nu} \propto \nu^{\alpha}$) and a turnover which is
relatively flat.  PG1440+356 shows a very flat optical$-$near-IR
continuum ($\alpha_{ir} \sim -1$), a strong cool component compared
with the low-redshift median~\cite{Elvis94} and a steeper turnover
($\alpha_{cut} \sim 1.8$) which, however, still remains below the
critical value of 2.5 indicating thermal emission.  The range of IR
continua in the sample overlaps but extends to redder continua than
an optical/radio selected sample~\cite{Elvis94}.  This is apparent in
the low mean for the near-IR slope, $\alpha_{IR} \sim -1.4 \pm 0.5$
compared with the Median SED value of $-1.0 \pm 0.3$ and also the
extension to cooler values of the ratio of decade luminosities:
L$(1-10\mu)$/L$(10-100\mu) = 0.8 \pm 0.5$ compared with the
radio/optical sample: $1.0 \pm 0.8$.  The latter is also illustrated by
comparison of the histograms in Figure~\ref{fg:hotnesshist}.  However a
Kolmogorov-Smirnov test does not indicate a significant difference in
the samples (P=18\%).  If real, this trend is an interesting extension
to the lack of bias against red optical/UV continua in these sources.
Confirmation awaits more systematic comparisons in an upcoming paper.

\section{High Redshift AGN}


While skewed to low-redshift objects, due in part to lost sensitivity
and changes in observing mode, the Key Project sample contains 8
quasars in the redshift range $1.0 \leq z \leq 4.7$.  In order to
cover the intrinsic FIR continuum in these sources, we also have an
on-going sub-mm observing program at the JCMT.

\begin{figure}[h]
\centering
\vskip -0.2in
\includegraphics[width=.8\textwidth]{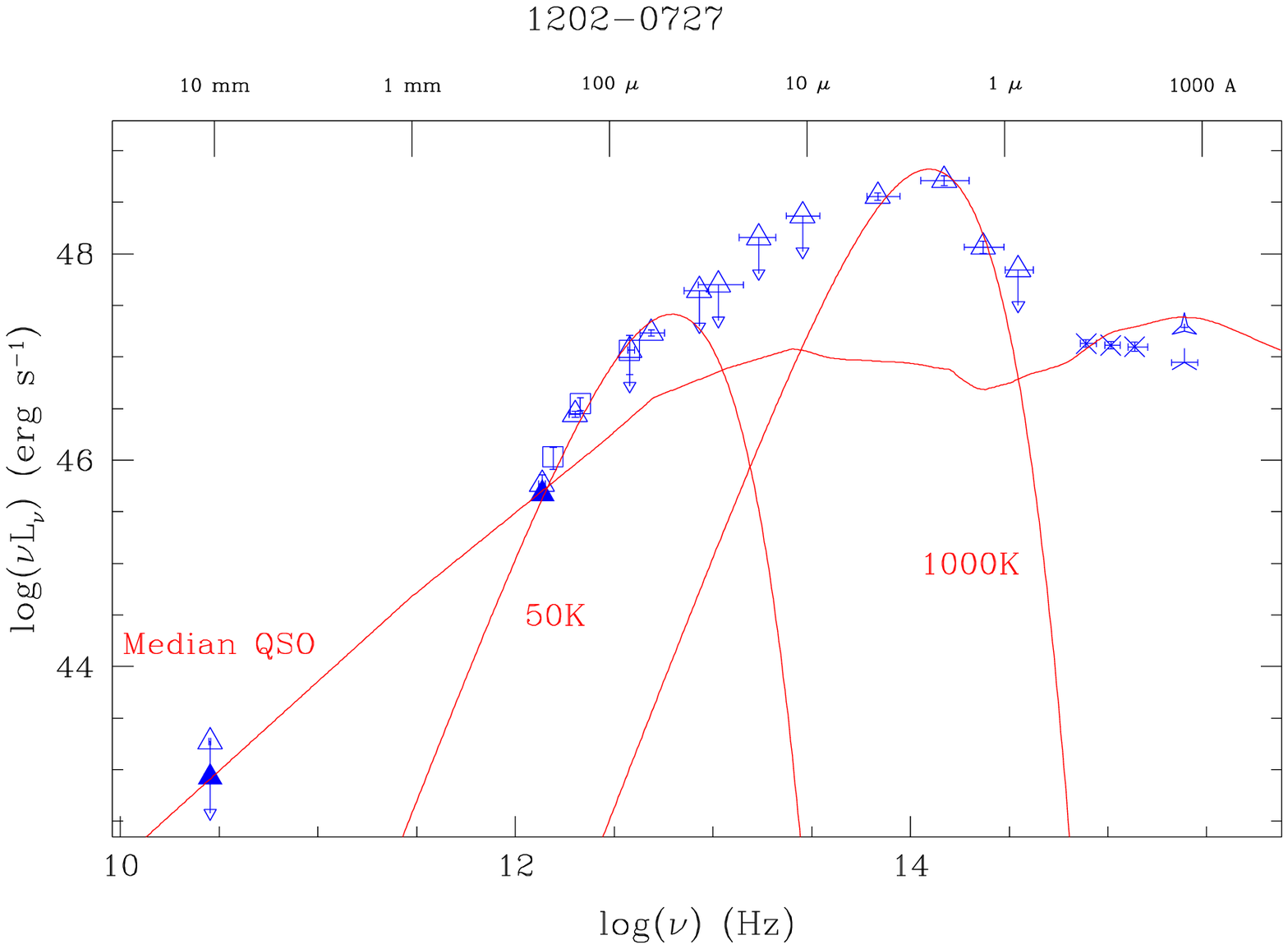}
\vskip -1.3in
\includegraphics[width=.8\textwidth]{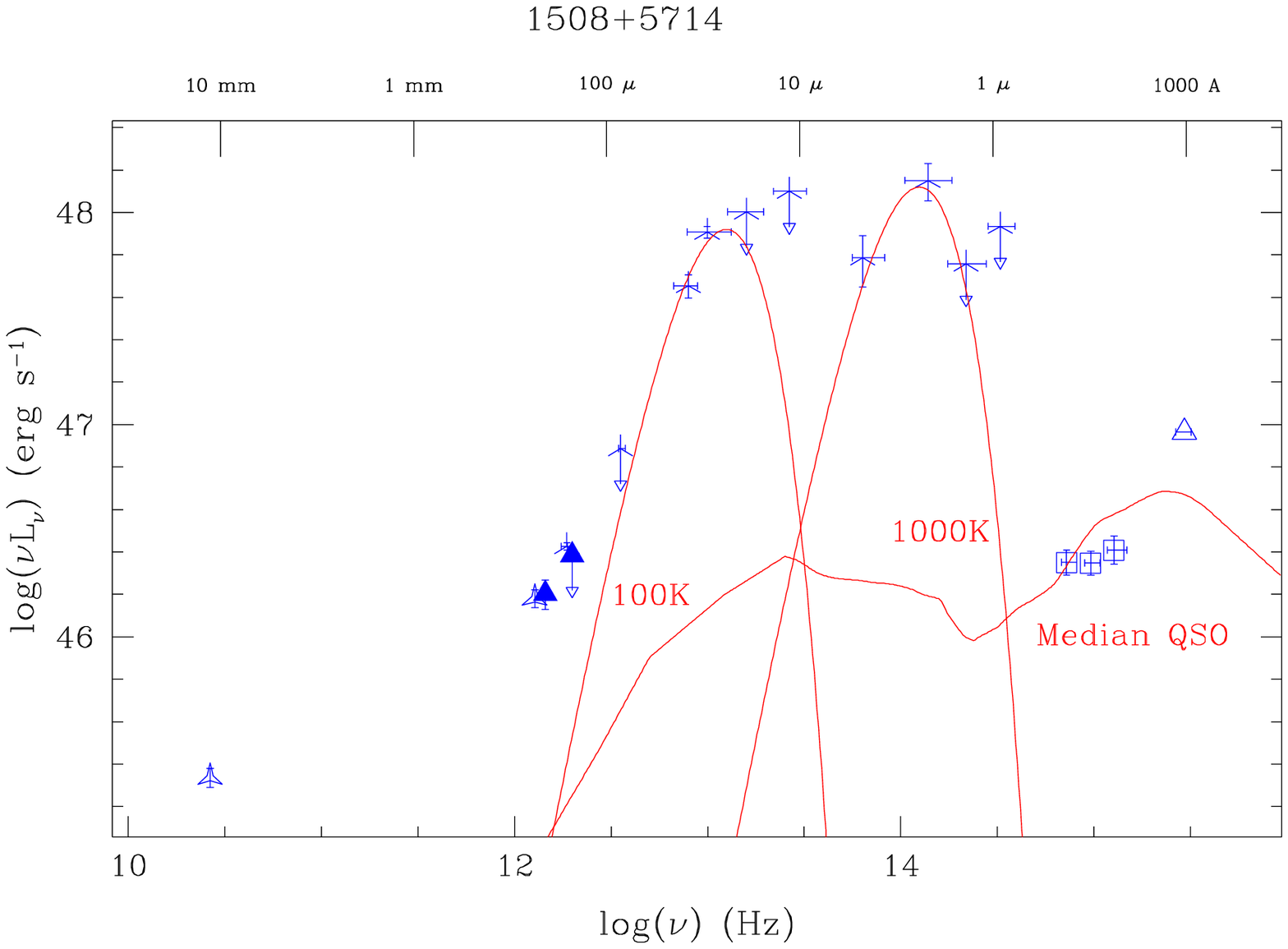}
\vskip -1.2in
\caption{\it Spectral Energy Distributions (SEDs) of a) 1202$-$0727 and b) 
1508+5714
compared with the median SED for a low-redshift sample~\cite{Elvis94}
illustrating the extremely strong near-IR emission in both sources.
Grey-body curves illustrating the minimum range of temperatures
for purely thermal emission are also shown.}
\label{fg:hizseds}
\vspace*{-0.2in}
\end{figure}

The spectral energy distributions of these quasars vary widely. The
highest redshift sources: 1202$-$0727 ($z = 4.69$); 1508+5714 ($z =
4.3$); and 1413+1143 ($z = 2.551$), are factors of 10--100 brighter in the
near-IR than
the low-$z$ median SED~\cite{Elvis94} (Figure~\ref{fg:hizseds}).  At
lower redshifts, PG1407+265 ($z = 0.944$) and PHL5200 ($z = 1.981$)
are a much closer match to the median, while PG1718+481 ($z = 1.084$)
is $> 5$ times fainter throughout the IR.  A much larger sample
will be required to determine the underlying distribution of SEDs at
moderate and high-$z$ which is hinted at by the variety seen here.

The strong IR emission in the highest redshift quasars places them
amongst the most powerful sources known. The range of temperatures is
broad, 40--1000 $K$.
Table~\ref{tb:hiz} lists derived properties for two sources.  The mass
estimates assume optically thin gray dust with $\beta = 2$.  The
$40\ts K$ component in 1508$+$5714 is based on only the observed
850$\ts \umu \rm{m}$ flux; it is likely to be an upper limit,
because the FIR slope is flattening at those wavelengths, indicating
possible contamination by a high frequency extension of the
non-thermal radio emission.  These and other high-redshift
AGN~\cite{Andreani99,Hughes93,McMahon99} contain large amounts of
dust, which implies a substantial star formation rate (SFR).  Average SFRs
listed in Table~\ref{tb:hiz} were calculated from the inferred mass assuming
a 1\% dust production efficiency~\cite{Blain99} and an onset of star
formation at $z = 10$.  Estimates of the FIR luminosity due solely to
this SFR, computed with a conversion of $2.2 \times 10^9 \ts L_
{\odot} \ts M_{\odot}^{-1} \ts {\rm yr}$~\cite{Rowan-Robinson97}, are
also listed.  These are $\sim$ factor 10 below the observed
luminosities, implying that the bulk of the FIR emission is driven by
the AGN, or the sources are undergoing an intense starburst of $\geq
1000 \ts M_{\odot} \ts {\rm yr}^{-1}$.

Both sources have large excesses over the median SED in a spectral
region that corresponds to warm and hot dust emission.  This emission
cannot yet be conclusively attributed to a thermal source, but that is
a strong possibility, since cool dust is clearly present in
1202$-$0727 and both objects show a sharp cutoff at wavelengths
corresponding to the Wien region of dust near the sublimation
temperature.  The difficulty with a dust interpretation is the energy
source to heat the material; the observed rest-frame optical and
near-UV luminosity is clearly too low in both sources.  Dust may
obscure all but a small fraction $\sim$ 1\% of the intrinsic blue bump
luminosity.  It is unlikely that we are directly viewing the extincted
source, as broad emission lines are clearly visible on continua which
do not appear heavily reddened~\cite{Storrie-Lombardi96}.  The
observed flux may be diminished by reflection off of a scattering
surface positioned above a dust torus.  Alternatively, if we are
seeing the intrinsic output of the central engine in the optical and
near-UV, then these AGN must be abnormally luminous in the far-UV and
X-rays.  There is an indication of a blue upturn in 1508$+$5714, but
the requisite luminosity remains unobserved.

\begin{table}
\centering
\vskip -0.2in
\caption{Properties of two High-Redshift Sources}
\label{tb:hiz}
\begin{tabular}{lllllll}
\hline

Name \rule{0cm}{3.5mm} &  $z$ & \multicolumn{3}{c}{Dust Mass
($M_{\odot}$)} \hspace*{1mm} & $\overline{\rm SFR}$ 
 & FIR($\overline{\rm SFR}$)  \\ 
& & $40 \ts K$ \hspace*{0.1in} & $100 \ts K$ \hspace*{0.1in} &
$1000 \ts K$ \hspace*{0.1in} & ($M_{\odot} \ts {\rm yr}^{-1}$)
\hspace*{1mm} & ($10^{45}$ erg s$^{-1}$) \\ 

\hline
1202$-$0727 \rule{0cm}{3.5mm} & 4.690 \hspace*{1mm} & $1.5 \times
10^9$ & $< 2.2 \times 10^7$  \hspace*{0.5mm} & 300  & 250 & 2.1 \\
%
%
1508+5714 \hspace*{1mm} & 4.300 & $2.0 \times 10^9$ \hspace*{1mm} & 
$3.6 \times 10^7$ & 53 & 280 & 2.4 \\
%
\hline
\end{tabular}
\vspace*{-0.4in}
\end{table}

\section{Summary}

The principal points discussed in this paper are:

\begin{itemize}

\item The X-ray selected AGN show a broader, cooler range of IR continua
than optical/radio selected AGN. 

\item The X-ray selected sample show no strong
evidence ((MR2251$-$178 excepted) for a steep, $\alpha_{cut} > 2.5$,
far-IR turnover, which would rule out non-thermal emission. 
Sub-mm observations are necessary to confirm.

\item $z > 4$ quasars show $\sim$100 times stronger near-IR emission than
low-$z$ AGN. This may be due to strong obscuration, orientation
or a real deficit of optical emission in these sources.

\item The strong cool IR emission in the high-$z$ sources implies dust
masses $> 10^9 \ts M_{\odot}$, which would require an average SFR of a
few hundred $M_{\odot} \ts {\rm yr}^{-1}$.  This underproduces the
observed FIR flux, so there is either a major AGN component, or the
SFR at the time of observation is significantly larger, $\geq 1000 \ts
M_{\odot} \ts {\rm yr}^{-1}$.

\end{itemize}

\clearpage
\addcontentsline{toc}{section}{Index}
\flushbottom
\printindex

\end{document}